\documentclass[aps,prl,twocolumn,showpacs,amsmath,amsmath,amssymb,amsfonts,longbibliography]{revtex4-1}
\usepackage[margin=0.8in]{geometry}
\usepackage{graphicx}
\usepackage{amsmath}
\usepackage{amssymb}
\usepackage{amsthm}
\usepackage{bbm}
\usepackage{mathrsfs}
\usepackage{float}
\usepackage{bm}
\usepackage{xcolor}
\usepackage{bbold} %This package allows to write the identity symbol
\usepackage{ifpdf}

\newcommand{\be}{\begin{equation}}
\newcommand{\ee}{\end{equation}}
\newcommand{\ba}{\begin{align}}
\newcommand{\ea}{\end{align}}
\newcommand{\sysb}{\left\{\begin{array}}
\newcommand{\syse}{\end{array}\right.}
\newcommand{\baa}{\begin{array}}
\newcommand{\eaa}{\end{array}}
\newcommand{\bs}{\begin{split}}
\newcommand{\es}{\end{split}}
\newcommand{\matb}{\left(\begin{array}}
\newcommand{\mate}{\end{array}\right)}

\newcommand{\rme}[1]{{\rm{e}}^{#1}}
\newcommand{\mand}{\quad\text{ and }\quad}

\newcommand{\id}{\mathbb{1}}

\newcommand{\ha}{\frac{1}{2}}

\newcommand{\lt}{\left(}
\newcommand{\rt}{\right)}

\newcommand{\lan}{\left\langle}
\newcommand{\ran}{\right\rangle}

\newcommand{\av}[1]{\lan #1 \ran}

\newcommand{\ket}[1]{\left| #1 \ran}
\newcommand{\bra}[1]{\lan #1 \right|}

\newcommand{\acomm}[2]{\left\{ #1, #2 \right\}}

\newcommand{\suml}[2]{\sum\limits_{#1}^{#2}}

\newcommand{\uar}{\uparrow}
\newcommand{\dar}{\downarrow}
\newcommand{\conf}{\mathcal{C}}

\newcommand{\BCount}{B}
\newcommand{\PCount}{\mathcal{P}}
\newcommand{\JumpCount}{q}

\newcommand{\NumOfSites}{L}
\newcommand{\NumOfExcitations}{N}
\newcommand{\NumOfSpinsFlipped}{s}

\newcommand{\HDiag}{H_C}
\newcommand{\HOffDiag}{H_Q}

\begin{document}
\title{Emergent kinetic constraints in open quantum systems}
\author{B. Everest}
\affiliation{School of Physics and Astronomy, University of Nottingham, Nottingham, NG7 2RD, United Kingdom}
\author{M. Marcuzzi}
\affiliation{School of Physics and Astronomy, University of Nottingham, Nottingham, NG7 2RD, United Kingdom}
\author{J.P. Garrahan}
\affiliation{School of Physics and Astronomy, University of Nottingham, Nottingham, NG7 2RD, United Kingdom}
\author{I. Lesanovsky}
\affiliation{School of Physics and Astronomy, University of Nottingham, Nottingham, NG7 2RD, United Kingdom}

\begin{abstract}
Kinetically constrained spin systems play an important role in understanding key properties of the dynamics of slowly relaxing materials, such as glasses. So far kinetic constraints have been introduced in idealised models aiming to capture specific dynamical properties of these systems. However, recently it has been experimentally shown by [M. Valado \textit{et al.}, arXiv:1508.04384 (2015)] that manifest kinetic constraints indeed govern the evolution of strongly interacting gases of highly excited atoms in a noisy environment. Motivated by this development we address and discuss the question concerning the type of kinetically constrained dynamics which can generally emerge in quantum spin systems subject to strong noise. We discuss an experimentally-realizable case which displays collective behavior, timescale separation and dynamical reducibility.
\end{abstract}

\maketitle

\emph{Introduction ---}
Understanding and characterizing the dynamics of strongly-interacting many-body systems remains a relevant challenge. This is even more the case in the context of systems undergoing complex collective relaxation such as glass formers which, under certain conditions (typically, below a certain temperature), display extremely long relaxation times \cite{YearsAging, ExpAging, Binder2005, Ritort2003, Peliti2012, Biroli2013, Berthier2016}. One approach proposed to explain this dynamical behavior assumes that on the microscopic level local transitions are only permitted when certain conditions, e.g. very specific arrangements of particles, are satisfied.
%focuses on the microscopic mechanisms for relaxation in terms of conditions for which local transitions are permitted.
These so-called ``kinetic constraints" \cite{Garrahan2002a,Ritort2003,Garrahan2007, Garrahan2009, Berthier2011} can produce dramatic effects on the dynamics: at sufficiently high densities or low temperatures there are severe restrictions on the allowed pathways that connect different many-body configurations. In practice this is achieved, e.g., via the energetic suppression of straightforward rearrangements. The remaining transitions then assume a highly cooperative character.

%, devising} forms of constraints which act through reciprocal hindrance of the elementary constituents or more sophisticated prescriptions. These \emph{kinetic constraints} \cite{Garrahan2002a,Ritort2003,Garrahan2007, Garrahan2009, Berthier2011} can produce dramatic effects on the dynamics; at sufficiently high densities/pressures and/or low temperatures, typical \changer{connected configurations} \mm{Why connected?} might become separated by high energy barriers, often corresponding to the requirement of extensive rearrangements of the system to pass from one to the other.

%A possible way to characterize this collective behavior is to introduce kinetic constraints in classical spin systems. \changer{One class of models are constrained lattice gases \cite{Ritort2003} where an excitation's hopping is constrained by its neighbors, mimicking excluded volume in dense fluids. Another are facilitation models, where} two paradigmatic examples are the so-called East \cite{EastMod} and Fredrickson-Andersen (FA) model \cite{Fredrickson1984}. Both \changer{of these examples} are stochastic processes which end up in \changer{a trivial and non-interacting steady state} ($\propto \rme{-\beta N}$, with $N$ the number of excited (up) spins and $\beta$ the inverse temperature), but feature dynamical rules which make the approach to stationarity highly intricate \changer{and often result in the emergence of meta-stability} \cite{Fredrickson1984, EastMod, Garrahan2007, Garrahan2009, Ritort2003}.

Depending on the specific mechanism, kinetically constrained models (KCMs) can be grouped into classes \cite{Ritort2003}. One set of examples are constrained (dynamic) lattice gases \cite{Kob1993, Teboul2014}, where a particle's diffusion (by hopping) is hindered by its neighbors, mimicking excluded volume in dense fluids. Another instance is given by facilitated spin models, such as the so-called East \cite{EastMod} and Fredrickson-Andersen (FA) models \cite{Fredrickson1984}, in which a spin's ability to update its state depends on the configuration of the ones nearby. Most KCMs consist of stochastic processes which end up in trivial and non-interacting steady states $ \rho_\mathrm{ss} $ (e.g., for the East and FA models, $ \rho_\mathrm{ss} \propto \rme{-\beta N}$ with $N$ the number of excited (up) spins and $\beta$ the inverse temperature), but feature dynamical rules which make the approach to stationarity highly intricate, often resulting in the emergence of meta-stability \cite{Fredrickson1984, EastMod, Garrahan2007, Garrahan2009, Ritort2003}.

%For example, the FA model does not allow the flipping of isolated excitations, i.e. $\downarrow\uparrow\downarrow\not\rightleftarrows\downarrow\downarrow\downarrow$, but requires a neighboring excitation to facilitate the spin-flip, e.g. $\downarrow\uparrow\uparrow\rightleftarrows\downarrow\downarrow\uparrow$. This simple rule slows down the dynamics at low temperatures when excited spins are rare and thus the constraint rarely admits spin-flips. Isolated excitations have to propagate through the system before they can meet and coalesce, leading to the emergence of meta-stability

\begin{figure}
\begin{center}
\includegraphics[width=\columnwidth]{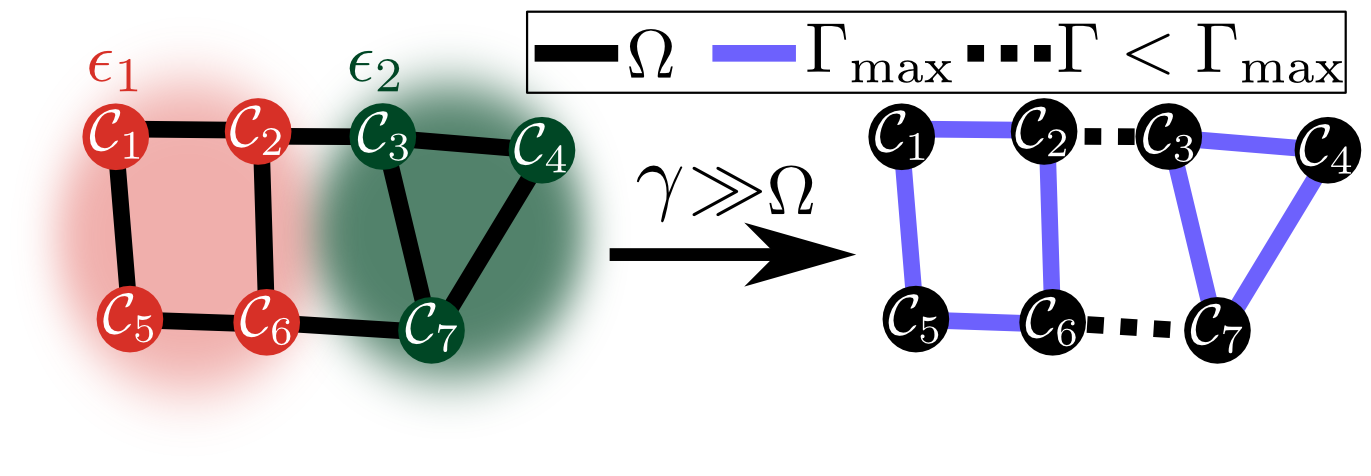}
\caption{Connectivity of the configuration space. Without noise (left), i.e., dephasing rate $\gamma = 0$, classical configurations, $\ket{\conf_m}$, shown as circles, are connected to each other by $\HOffDiag$ with coupling strength $\propto \Omega$. In this example $\HDiag$ is constructed such that the energy landscape in configuration space is separated into two plateaus with energies $\epsilon_1$ (red) and $\epsilon_2$ (green). This choice leads, in the presence of strong noise (right hand side), to two weakly connected spaces. The transition rates within and between the domains are $\Gamma$ and $\Gamma_\mathrm{max}$, respectively. For $\Gamma_\mathrm{max} \gg \Gamma$ this results in an (approximate) ergodicity breaking. For further explanation see text.
}
\label{fig:spaceCartoon}
\end{center}
\end{figure}

Despite their success in capturing hierarchical relaxation, it is only very rarely possible to derive kinetic constraints from first principles and they appear to remain an effective construct \cite{Keys2011}. However, it was recently shown that they naturally emerge in quantum optical systems, specifically cold atomic gases, in the presence of strong interactions and dephasing noise \cite{Lesanovsky2013b, Poletti2013}. In certain regimes, these systems show aspects of the facilitation dynamics \cite{Marcuzzi15, Everest15} inherent to the FA and East model, as highlighted in recent experiments \cite{Valado2015, Urvoy2015}. Kinetic constraints moreover govern the non-equilibrium dynamics of nuclear ensembles undergoing so-called Dynamic Nuclear Polarization \cite{Kockenberger2015} --- a process used to enhance the signal in magnetic resonance imaging applications. Further to that, a connection between kinetically constrained models and many-body localization in the absence of disorder was also established \cite{Hickey2014, VanHorssen2015}.

The aim of this work is to explore kinetic constraints that emerge in noisy quantum systems from a more general perspective. We discuss the construction of kinetically constrained models (KCMs),
and report an example of an effective reaction-diffusion process.
This experimentally realizable case displays pronounced collective behavior, timescale separation as well as dynamical reducibility of the state space --- features that are typically present in glassy dynamics. We also comment on the realizability, within our approach, of prototypical glass models such as the FA and East models.

%
% analyzing the form of the emerging constraint functions. We show that due to the \changer{structure} \mm{I would not use the term topology, since most physicists nowadays associate it to simple-connectedness only} of the state space and the fact that --- owed to our construction --- the stationary state is typically an infinite temperature state, canonical glass models (FA and East model) \changer{cannot} be realised. Nonetheless, systems with highly intricate relaxation dynamics can still emerge as a consequence of the kinetic constraints introducing conserved quantities, which results in ergodicity breaking.

\emph{Construction of kinetically constrained spin systems ---}
We focus here on spin$-\ha$ systems (with internal states $\ket{\uar}$, $\ket{\dar}$) arranged on a regular lattice --- whose $L$ sites are labeled by an index $k$ --- with the standard spin operators $\sigma_k^+ = \ket{\uparrow_k}\bra{\downarrow_k}$, $\sigma_k^- = (\sigma_k^+)^{\dagger}$, $\sigma_k^z = [\sigma_k^+,\sigma_k^-]$. The coherent evolution of the spins is governed by a Hamiltonian $H = H_C + H_Q$ which we separate into a ``classical'' and a ``quantum'' part. The former assumes the form
\begin{align}
H_C= & \sum_k u_k n_k + \sum_{k,j} \frac{v_{kj}}{2} n_k n_j + \sum_{k,j,i} \frac{ w_{kjl} }{3!} n_k n_j n_l + ...,
%&\frac{1}{3!} \sum_{k,j,i} w_{kjl} n_k n_j n_l + ...,
\end{align}
%where $n_k\ket{\uparrow} = 1$ and $n_k\ket{\downarrow} = 0$, $E_k$, $V_{kj}$ and $W_{kji}$ are the one, two, three body interactions respectively. This acts to produce the energies of different configurations, $H_C\ket{\mathcal{C}_m} = E_m\ket{\mathcal{C}_m}$, as displayed for two regions of different energies on the L.H.S of Fig. \ref{fig:spaceCartoon}.
where $n_k = (\sigma_k^z + 1)/2 = \ket{\uar_k} \bra{\uar_k} $ and $u_k$, $v_{kj}$, $w_{kjl}$ can be interpreted as one-, two-, three-body interaction couplings. This part defines an energetic landscape $E_m$ over the classical (Fock) configurations $\ket{\conf_m} = \ket{\cdots \uar_{k-1} \, \uar_{k}\, \dar_{k+1} \cdots}$ ($m = 1 \ldots 2^L$) via $ H_C \ket{\conf_m} = E_m \ket{\conf_m}$. In Fig.~\ref{fig:spaceCartoon}, these configurations are represented as circles and grouped in domains of equal energy.
%each is associated to a corresponding classical energy $E_m$ ($ H_C \ket{\conf_m} = E_m \ket{\conf_m}$).

The quantum part acts as $H_Q \ket{\conf_m} = \sum_{n \neq m} a_{mn} \ket{\conf_n}$ and defines the dynamical connectivity of the configurations. This is illustrated in Fig.~\ref{fig:spaceCartoon} where the solid lines correspond to the cases in which $a_{mn} \neq 0$. Here we focus on two prototypical examples:
% defines the connectivity of the states, shown as the lines in Fig. \ref{fig:spaceCartoon}, where we pick out two cases.
Spin-flipping, induced e.g.~by a laser on a two-level atomic transition which is commonly implemented in Rydberg atomic systems \cite{Saffman2010, Low2012, Labuhn2015}; and quantum tunneling of hard-core bosons between nearest neighbors \cite{Bloch2008a, Garcia-Ripoll2009a}, which are described by the Hamiltonians
\be
	 \HOffDiag^{(f)} = \Omega  \suml{k}{}  \sigma_k^x \mand \HOffDiag^{(t)} = \Omega  \suml{\av{k,j}}{}  \sigma_k^- \sigma_j^+   , \label{eq:offDiagonalHamiltonians}
\ee
respectively. Here $\av{k,j}$ is shorthand for summing over nearest neighbors only, $\sigma_k^x = \sigma_k^+ + \sigma_k^-$, and $\Omega$ is the coupling strength of the two processes (i.e., depending on the realization: the laser Rabi frequency, the exchange coupling or lattice tunneling amplitude).

The system is in contact with an environment which induces fast decoherence of quantum superpositions. We assume the noise to be white and spatially uncorrelated, so that the evolution of the density matrix $\rho$ is governed by the Lindblad equation \cite{Lindblad1976, Breuer2002} 
%$\dot{\rho} = -i[H,\rho] + \gamma\sum_k (n_k\rho n_k -\frac{1}{2}\lbrace n_k,\rho \rbrace )  $,
%
\be
\dot{\rho} = -i[H,\rho] + \gamma\sum_k^\NumOfSites(n_k\rho n_k -\frac{1}{2}\lbrace n_k,\rho \rbrace )  
%\equiv \mal{L} \rho, 
\label{eqn:masterEqn}\\
\ee
where $\acomm{A}{B} = AB+BA $ denotes anticommutation and $\gamma$ is the dephasing rate. This form of dissipation occurs naturally in cold atom lattice experiments, stemming e.g.~from the off-resonant scattering of photons from the optical-trapping laser field \cite{Sarkar2014a}, or from phase noise of the laser driving \cite{Urvoy2015, Valado2015, Schempp2014}.
We further set $\gamma \gg \Omega$. This allows the adiabatic elimination of $\HOffDiag$ and the projection of the dynamics onto the subspace of diagonal density matrices $\mu$ in the $\ket{\mathcal{C}_m}$ basis \cite{Lesanovsky2013b, Barthel2013, Marcuzzi14, Degenfeld14, Bernier2014a, Sciolla2014}. The reduced state $\mu$ can then be interpreted as a probability distribution and evolves according to the classical master equation
\begin{align}
\partial_t \mu =  \sum_\nu \frac{4}{\NumOfSpinsFlipped \gamma} \Gamma_\nu \left( l_\nu^{\dagger} \mu l_\nu + l_\nu \mu l_\nu^{\dagger} - \lbrace l_\nu, l_\nu^{\dagger} \rbrace \mu  \right). \label{eqn:classicalMasterEquation}
\end{align}
where the operators $l_\nu$, the index $\nu$ and the coefficient $s$ depend on the choice of $H_Q$. In the case of spin flipping $H_Q^{(f)}$ the sum runs over the sites $\nu \equiv k$, $s = 1$ and $l_k = \Omega \sigma_k^+ $. For tunneling, instead, the sum runs over neighboring pairs $\nu \equiv \av{kj}$, $s = 2$ and $l_{kj} = \Omega \sigma_k^+ \sigma_{j}^-/\sqrt{2}$.
%
%Note that the natural steady state of Eq. \eqref{eqn:classicalMasterEquation} is proportional to the identity $\mu_\mathrm{ss} \propto \id$ (for the moment we discard the possibility of ergodicity breaking which we will discuss later). Despite this trivial stationary state non-trivial dynamics stem from the configuration-dependent rates $\Gamma_k$ which give rise to kinetic constraints. 
The rates $\Gamma_\nu$ are configuration-dependent and read
\begin{align}
 \frac{1}{\Gamma_\nu} = 1 + \left( \frac{2\,\delta E }{\NumOfSpinsFlipped \gamma} \right)^2, \label{eqn:rate}
\end{align}
where $\delta E$ is the ``energy cost'' of performing the $l_\nu$-induced transition. More precisely, when $l_\nu \ket{\conf_m} \propto \ket{\conf_n}$ then $\delta E = E_n - E_m$. Note that the inverse process induced by $l_\nu^\dag$ occurs at the same rate; therefore, Eq.~\eqref{eqn:classicalMasterEquation} satisfies detailed balance at infinite temperature and the steady-state distribution $\mu_{\mathrm{ss}}$ is uniform ($\propto \id$ under ergodic conditions).

\emph{``Hard" and ``soft" kinetically constrained models ---}
According to (\ref{eqn:rate}) the rate of a transition is maximal when both involved states are on resonance, i.e. $\delta E = 0$. Conversely, if $|\delta E| \gg \gamma$ the transition rate is greatly suppressed. This implies that depending on the precise form of $\HDiag$, particular processes can be favored over others, thereby constraining in turn the dynamics to favor specific pathways in configuration space.

%Hence, by tuning the form of $\HDiag$ one can favor the occurrence of some processes over some others, thereby constraining the dynamics to favor specific pathways in configuration space.
%We argue here that kinetic constraints can be imposed by solely acting on the properties of $\HDiag$ when coupled with one of the two $\HOffDiag$'s.

In the limit $|\delta E| / \gamma \to \infty$ the suppression is total and the corresponding transition is ``blocked''. Ideally, in a context where energy differences are either vanishing or infinite, one obtains a \emph{hard} constraint and transitions induced by $H_Q$ either take place at rate $\Gamma_{\mathrm{max}} = 1$ or never occur ($\Gamma = 0$). As highlighted in Fig.~\ref{fig:spaceCartoon}, this causes the space to fragment into disconnected parts (corresponding to different energies), breaking ergodicity and producing a reducible dynamics. Necessarily, any kinetic constraint prohibiting a transition between two configurations ($\ket{\conf_1} \not\to \ket{\conf_2}$) can only admit a hard realization if these belong to dynamically-disconnected sub-spaces, i.e., if there is no sequence of allowed transitions connecting them.
%\changer{This makes the effective dynamics reducible.}

If such a pathway exists, (e.g., $\ket{\conf_1}\rightarrow\ket{\conf_3}\rightarrow\ket{\conf_4}\rightarrow\ket{\conf_2}$), the realization of a \emph{soft} constraint \cite{Elmatad2010} might still be possible. In this case direct transitions between $\ket{\conf_1}$ and $\ket{\conf_2}$ cannot be forbidden but merely suppressed. The degree of suppression is determined by the minimal number $q$ of allowed transitions joining $\ket{\conf_1}$ and $\ket{\conf_2}$ and is $\Gamma_{\mathrm{suppressed}} / \Gamma_{\mathrm{allowed}} \gtrsim 1/\JumpCount^2$. 
%If there are $\JumpCount$ connecting transitions, each with fixed energy difference $\overline{\delta E}$, the suppressed transition will feature an energy difference $\JumpCount \times \overline{\delta E}$ leading, if $\gamma \ll \overline{\delta E} < \infty$, to a ratio $\Gamma_{\mathrm{suppressed}} / \Gamma_{\mathrm{allowed}} \approx 1/\JumpCount^2$. This case is extremal, i.e., even for a non-uniform choice of the $\delta E$s, the minimal ratio cannot be made smaller than this. 

%The situation does not improve by allowing for a non-uniform choice of the $\delta E$s on the intermediate steps.
%If one allows for the generation of other suppressed transitions that do not lie within the constrained model then the ratio may be improved.

%without generating other transitions which are more suppressed than others, which may not conform to the constraint.
%Determining whether a soft realization is possible and to what extent it can reproduce the constraint is quite involved as it introduces a concept of directionality (related to the sign of $\delta E$ on each link in Fig.~\ref{fig:spaceCartoon}) and is beyond the scope of this work.

\begin{figure}
\begin{center}
\includegraphics[width=\columnwidth]{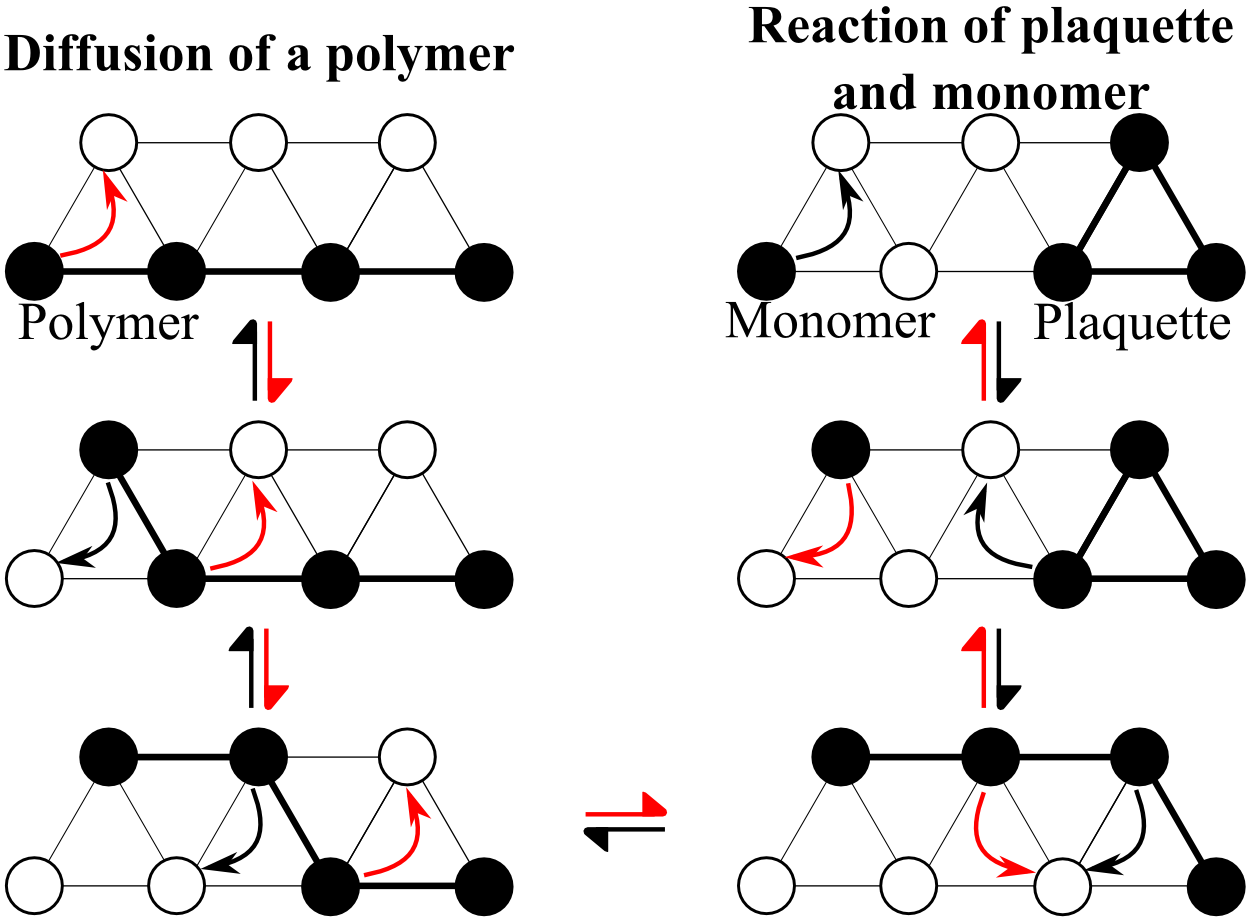}
\caption{Illustrations of key processes governing the dynamics of the reaction diffusion model with constant bonds. The black dots represent excitations and a thick line on the lattice represents a bond. Arrows denote possible moves and point at the resulting configurations. A polymer (left) is a chain of connected excitations, and shown is the way in which it can diffuse across the lattice. A plaquette (right) is formed by three excitations filling the vertices of a triangular tile. We showcase its reaction with a monomer.}
\label{fig:polymerCartoons}
\end{center}
\end{figure}

\emph{Reaction-diffusion model with constant bonds ---}
Based on the above discussion, we construct here a KCM which
%, in which hopping of excitations is dependent upon the nearby configuration, 
mimics a lattice gas with excluded volume effects. This model admits a hard realization and is simple enough to be experimentally realizable with cold atoms in an optical lattice (see Refs.~\cite{Sarkar2014a, Dutta2015}). It consists of particles arranged on a triangular lattice which feature nearest neighbor tunneling, as given by $\HOffDiag^{(t)}$ in Eq.~\eqref{eq:offDiagonalHamiltonians}, and strong nearest-neighbor interactions, $\HDiag = U \sum_{\langle k, j \rangle} n_k n_j$. In the presence of dephasing this leads to a stochastic process of excitations (up spins) hopping with rates that depend on the interaction strength $U$. By construction the number of excitations $\NumOfExcitations$ is conserved. Taking the limit $U / \gamma \to \infty$ introduces a further conserved quantity, namely the number $B$ of neighboring pairs of excitations (bonds). Consequently, excitations can only hop if doing so preserves the number of bonds between them. 

Clusters of excitations become bound structures, whose dynamical behavior strongly depends on their shape. Two primary examples are shown in Fig.~\ref{fig:polymerCartoons}. The first is a ``polymer", consisting of two or more excitations arranged along a chain, which can only diffuse via slow, cooperative motion \cite{Jones2011}. The second is a ``plaquette'', three excitations at the vertices of the same triangular tile. The plaquette is the simplest example of an immobile structure which cannot diffuse by itself, since any hop would result in the net loss of (at least) a bond. It can, however, react with ``monomers'' (isolated excitations) or other mobile structures (see r.h.s.~of Fig.~\ref{fig:polymerCartoons}). This leads to an assisted diffusion which is reminiscent of the strongly cooperative motion found in many glassy models \cite{Fredrickson1984, EastMod, Garrahan2007, Garrahan2009, Ritort2003}.

\begin{figure}
\begin{center}
\includegraphics[width=\columnwidth]{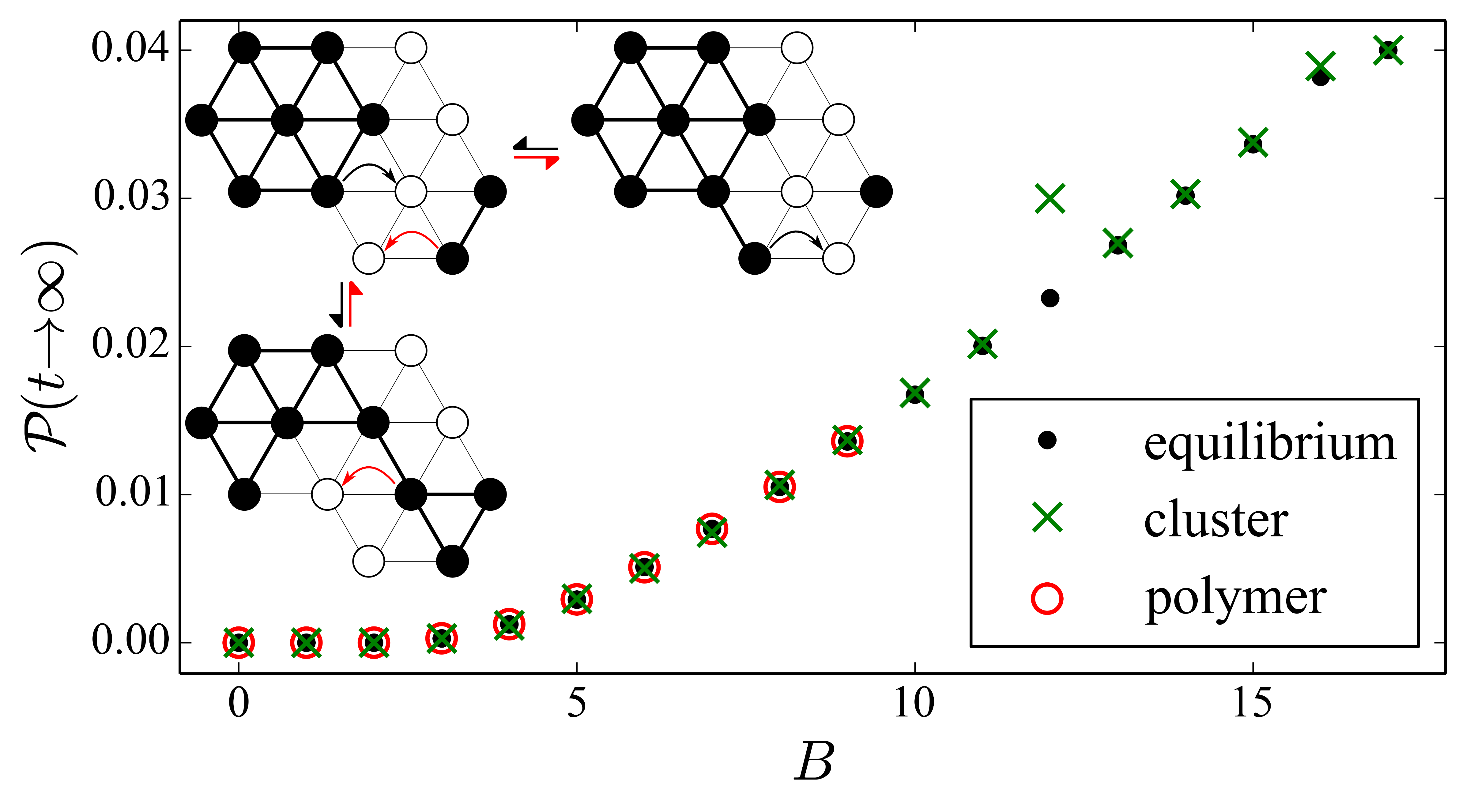}
\caption{Stationary plaquette density $\PCount (t \to \infty)$ against the number of bonds $B$ for a $10 \times 10$ lattice with $\NumOfExcitations = 10$, obtained via two different averaging procedures: black dots are calculated from a uniform random sampling of configurations at fixed ($N,B$). The remaining data points are averages over different realizations of the dynamics via a kinetic Monte Carlo procedure, differing by the initial condition: (green crosses) A single cluster plus monomers; the former is chosen to be as compact as possible. (red circles) A single polymer plus monomers. 
%The ``cluster" (green crosses) initial condition starts the excitations in as compact a configuration as possible. The ``polymer" (red circles) initial condition starts the excitations in a chain. 
The inset shows how the cluster present at $\BCount = 13$ can react with a monomer and the one at $\BCount = 12$ with a dimer.
%
%The first hinges on our knowledge of the fact that detailed balance holds at infinite temperature, and is an ``equilibrium'' method (black dots) and performs a uniform random sampling of configurations at fixed $N$ and $B$. The second initialises the system in a specific initial condition and evolves it via a kinetic Monte Carlo procedure, recording the final value of $\PCount$. The ``cluster" (green crosses) initial condition starts the excitations in as compact a configuration as possible. The ``polymer" (red circles) initial condition starts the excitations in a chain. The inset shows how the cluster present at $\BCount = 13$ can react with a monomer.}
}
\label{fig:equilibrium}
\end{center}
\end{figure}

%
%--------------
%
%Owed to the fact that 
%
%Interestingly, the system possesses apart from $\NumOfExcitations$ and $\BCount$ additional conservation laws. 

Interestingly, $N$ and $B$ do not exhaust all the conservation laws of this model. There are additional, subtler ones that further split the space of configurations. The easiest way to realize this is to consider the case $\NumOfExcitations = \BCount = 3$, which encompasses all possibilities of placing a single plaquette in the lattice: since plaquettes are unable to move on their own, all these states are dynamically disconnected. This finer structure is generally related to the formation of immobile clusters and thus emerges at high numbers of bonds $\BCount \gtrsim \NumOfExcitations$. This is exemplified in Fig.~\ref{fig:equilibrium}, where we compare results obtained from dynamical simulations with estimates based upon assuming that the steady state is an equilibrium ``microcanonical shell'' at fixed $(N,B)$. Without this additional dynamical reduction, the two predictions would coincide. Shown is the plaquette density $\PCount = (\text{\# of plaquettes}/2\NumOfSites)$ in a $10\times 10$ lattice with $\NumOfExcitations=10$ excitations and different values of $B$ from $0$ to $17$. The black dots are averages obtained from uniform random samplings of states at fixed $(N,B)$. The other data sets correspond to long-time values of $\PCount$ extracted from kinetic Monte Carlo simulations of the dynamics. The initial conditions are chosen either to have all bonds taken by a single polymer structure (red circles) -- which is only possible up to $\BCount = \NumOfExcitations-1 = 9$ -- or to have all bonds taken by the smallest possible cluster (green crosses). In both cases, the remaining excitations are introduced as monomers.

At sufficiently low number of bonds $B$ there are no appreciable deviations
%; furthermore, the result does not depend on the choice of the initial condition, which shows that 
and most configurations with the same $(N,B)$ are dynamically connected. 
%This breaks down at higher values: 
For $\BCount = 12$ and $16$, however, the ``cluster initialization'' displays a higher stationary plaquette density than the naive equilibrium value. For instance, the initial cluster at $\BCount = 12$ is chosen to be the ``filled hexagon'' displayed in the top-left corner of Fig.~\ref{fig:equilibrium}. Monomers cannot react with it, since each of the outer excitations forms three bonds. In order to break it apart, the assistance of a dimer (or longer polymer) is required. Therefore, for $\BCount = 12$ this structure is inert, while the remaining monomers explore the rest of the lattice via ordinary diffusion. Note however that adding bonds does not necessarily make a structure less prone to dissolution: for $\BCount = 13$ the initial cluster can react with monomers via the mechanism displayed in Fig.~\ref{fig:equilibrium}, starting from the top-right configuration.

% meaning that $(N,B)$ still provide enough information to identify the typical configurations explored by the dynamics. At values $\BCount = 12$ and $16$ the cluster state shows a higher stationary plaquette density than the naive equilibrium value. This uncovers the presence of further fragmentation of the configuration space; for instance, the starting cluster at $\BCount = 12$ is the ``filled hexagon'' displayed in the top-left corner of Fig.~\ref{fig:equilibrium}. Monomers cannot react with it, since each of the outer excitations forms three bonds; in order to break it apart even partially, the assistance of a dimer (or longer polymer) is required. Therefore, for $\BCount = 12$ this structure is immobile, while the remaining monomers explore the rest of the lattice without ever reacting. The case $\BCount = 13$ admits instead partial dissolution of the initial cluster; the mechanism is reported in Fig.~\ref{fig:equilibrium}, starting from the top-right configuration.

\begin{figure}
\begin{center}
\includegraphics[width=\columnwidth]{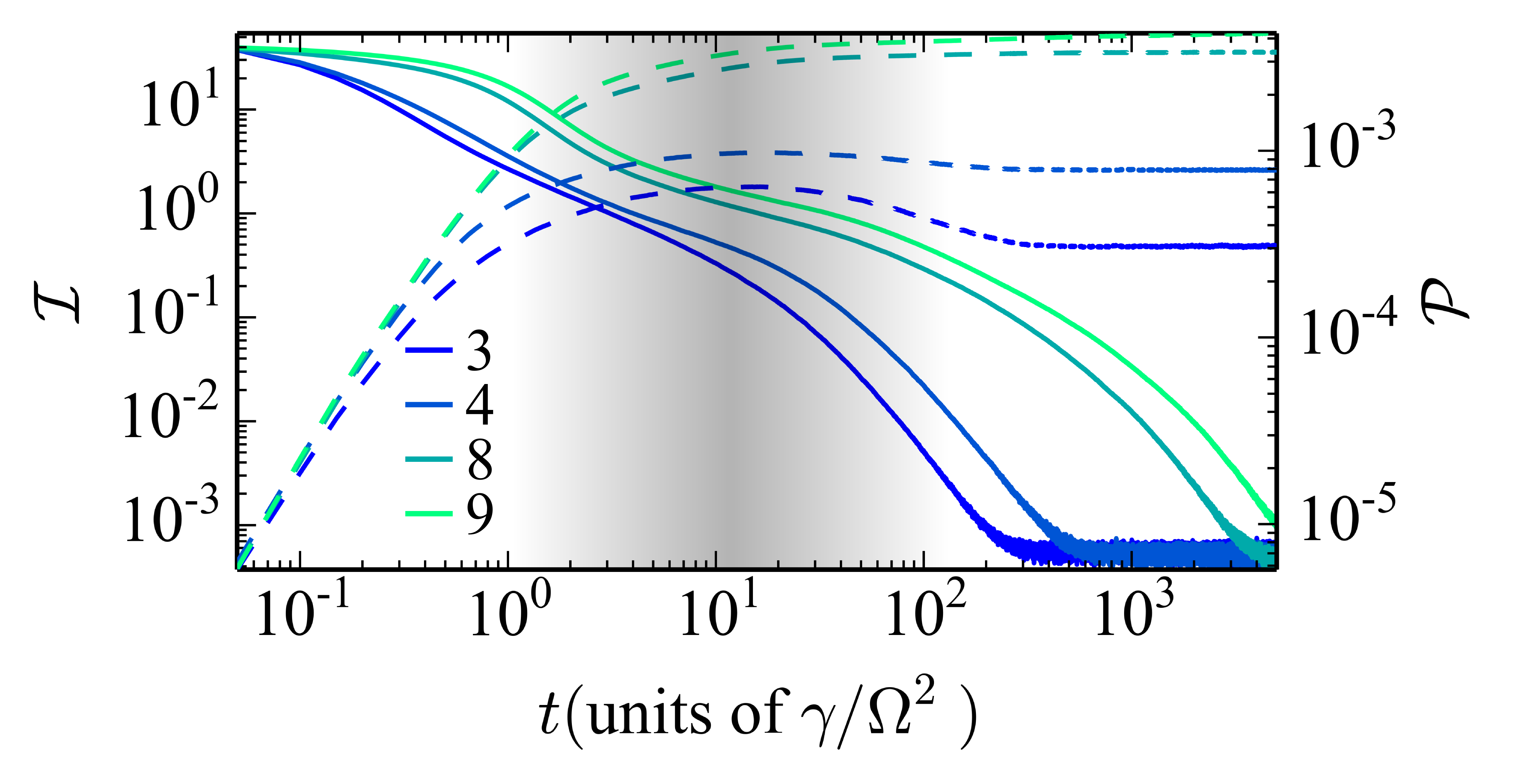}
\caption{Dynamics of the imbalance $\mathcal{I}$ (solid) and plaquette density $\PCount$ (dashed) for a $20\times 20$ lattice with $\NumOfExcitations = 10$. All cases are initialised with a single $(\BCount + 1)$ polymer and $\NumOfExcitations - \BCount -1$ monomers for $\BCount = 3$, $4$, $8$ and $9$.
% Four polymer initial conditions are shown with $\BCount = 3$, $4$, $8$ and $9$. 
The relaxation time increases with $\BCount$. At low $\BCount$, the plaquette density overshoots its stationary value and correspondingly the decay of the imbalance speeds up. This highlights the advantage in liberating monomers (or smaller structure) by forming plaquettes. The subsequent assisted diffusion of plaquettes acts on much longer timescales and eventually reduces $\PCount$ to its stationary value. 
%
%
%
%It highlights that although a plaquette is fixed in space, it also frees a smaller scale polymer in its production allowing faster spreading across the lattice. 
The shaded area marks the separation between two regimes in the dynamics, the earliest dominated by plaquette creation and monomer diffusion, the latest by assisted diffusion of plaquettes.}
\label{fig:combo}
\end{center}
\end{figure}

The presence of complex structures which cannot move by themselves and can only undergo assisted diffusion results in a separation of timescales in the dynamics, as displayed in Fig.~\ref{fig:combo}. There we report the evolution of the \emph{imbalance} $\mathcal{I}=\sum_{\langle k, j \rangle} |\langle n_k \rangle - \langle n_j \rangle|^2$, a measure of the non-uniformity of the system, and the plaquette density $\PCount(t)$ as a function of time for a $20\times 20$ lattice, $\NumOfExcitations = 10$, and prepared at $t=0$ in a single-polymer state with $\BCount = 3$, $4$, $8$ and $9$. These configurations are able to explore the entire lattice and thus to restore translational invariance for sufficiently long times, implying $\mathcal{I} (t \to \infty) \to 0$. The early dynamics is dominated by diffusion of the original structures (predominantly monomers) and formation of plaquettes. For the low-$B$ cases, around $t \approx 10 \gamma / \Omega^2 $  the plaquette density reaches its maximum, which is higher than its stationary value. Correspondingly, the imbalance relaxation speeds up, which can be understood as follows: 
%clusters offer a higher bond-to-excitation ratio than polymers; in other words, 
the formation of clusters such as plaquettes breaks down polymers to shorter ones, which display higher mobility and diffuse faster. For instance, for $\BCount = 3$, once a plaquette is formed an additional monomer is released (see Fig.~\ref{fig:polymerCartoons}) and monomers are the most efficient objects at exploring the lattice. Consequently, the higher the plaquette density, the higher the rate of relaxation of the imbalance. On longer time scales, further plaquette-monomer reactions relax $\PCount$ to its actual stationary value.

%
%ll the other excitations are monomers, which are the most efficient objects at exploring the lattice. As a result, once the plaquette density is maximized it leads to an increased rate of relaxation of the imbalance. After this, the plaquette density relaxes to its actual stationary value.

\emph{Facilitated spin models ---}
For completeness, we comment here on the realizability of the aforementioned (one dimensional) FA and East models \cite{Ritort2003}.
%mentioned above which play a prominent role in the understanding of systems that have a trivial thermodynamics, but display rich collective behavior in their relaxation towards stationarity \cite{Ritort2003}. 
Both feature facilitated spin flipping [$\HOffDiag^{(f)}$ in Eq.~\eqref{eq:offDiagonalHamiltonians}], whereby an excitation (up spin) enables the flipping of its neighbors e.g. $\uar \uar \dar \rightleftarrows \uar \dar \dar$ (whereas $\downarrow\uparrow\downarrow\not\rightleftarrows\downarrow\downarrow\downarrow$).
%\changer{We briefly investigate two KCMs, defined earlier, the FA \cite{Fredrickson1984} and the East model \cite{EastMod} which are both one-dimensional toy models for capturing glassy behavior. Both feature a facilitated spin flipping, whereby an excitation enables the flipping of its neighbors e.g. $\downarrow\uparrow\uparrow\rightleftarrows\downarrow\downarrow\uparrow$.  }
In the East model, facilitation is further constrained and can only take place to an excitation's right. Neither model admits a hard realization. To see this we consider the transition $\uar \dar \dar \dar \to \uar \dar \uar \dar$ which must be forbidden in both models. However, both configurations can be connected via a sequence of allowed steps $\uar \dar \dar \dar \to \uar \uar \dar \dar \to \uar \uar \uar \dar \to \uar \dar \uar \dar$.
% Note that a single spin-flip changes the parity of the number of excitations, which implies that any pair joined via a blocked transition can only be connected via an odd number of allowed transitions. We have therefore also determined that, in a soft realization, the allowed transitions will occur at best $3^2 = 9$ times faster than the non-allowed ones, making the constraint quite weak. Furthermore, by direct analysis of the configuration space one can show that no soft realization is possible for the East model, whereas the prescription $\HDiag = U \sum_k n_k \lt 1 - 2n_{k+1}/3 \rt$ works for FA, yielding $\delta E = \pm U/3$ in the presence of at least an excited neighbor and $\delta E = U$ otherwise.
The FA model still admits a soft realization (choosing $\HDiag = U \sum_k n_k \lt 1 - 2n_{k+1}/3 \rt$) with $\Gamma_{\mathrm{suppressed}} / \Gamma_{\mathrm{allowed}} \gtrsim 1/9$.

%As for soft realizations, one can show that only the FA model admits one, corresponding to $\HDiag = U \sum_k n_k \lt 1 - 2n_{k+1}/3 \rt$ with $\Gamma_{\mathrm{suppressed}} / \Gamma_{\mathrm{allowed}} \gtrsim 1/9$.

Furthermore, for the facilitated dynamics inherent to the FA and East models to display glassy features, it is crucial that the density of excitations (up spins) remain low. Conversely, under Eq.~\eqref{eqn:classicalMasterEquation} the state invariably evolves towards equilibrium at infinite temperature, which poses a severe restriction to its applicability in this case. However, introducing additional noise sources might provide a way around, as it may change the nature of the stationary state (see Refs.~\cite{Lesanovsky2013, Hoening2014, schempp2014full, Schonleber2014, Everest15}).

\emph{Conclusions ---}
Kinetically constrained models were originally introduced to capture the basic properties of slow-relaxing materials, yet have largely remained an idealized construct. Here we have shown that in the presence of strong noise these constraints emerge rather naturally in the dynamics of open quantum systems. As an example we discussed an experimentally realizable reaction-diffusion model which displays cooperative, reducible dynamics and we have highlighted the emergence of assisted diffusion processes which lead to timescale separation.

%, ergodicity breaking and timescale separation due to the emergence of assisted diffusion. 

The construction employed in this work results in effectively classical models. An interesting question is how the behavior of those changes when quantum coherence is not entirely washed out by the noise. This could be systematically addressed in an experimental realization of the discussed reaction-diffusion model with cold atoms in lattices \cite{Bakr2009, Bakr2010, Sherson2010} thereby providing a handle for exploring quantum effects in glassy relaxation \cite{Markland2011, Olmos2012}. This could also shed light on the interplay between quantum and classical fluctuations on collective phenomena, as e.g. recently discussed in \cite{Marcuzzi2016}.

\acknowledgements

The research leading to these results has received funding from the European Research Council under the European Union's Seventh Framework Programme (FP/2007-2013) ERC Grant Agreement No. 335266 (ESCQUMA), the EU-FET grant No. 612862 (HAIRS) and the H2020-FETPROACT-2014 Grant No. 640378 (RYSQ). We also acknowledge financial support from EPSRC Grant No. EP/M014266/1. Our work has benefited from the computational resources and assistance provided by the University of Nottingham High Performance Computing service.

\bibliography{kineticConstraints}

\end{document}